%% file: draft.tex
\pdfoutput=1
\documentclass[aps,prb,superscriptaddress,papersize=a4paper,twocolumn,10pt]{revtex4-1}
\usepackage{amsmath,amsfonts,amssymb,amsthm}
\usepackage{mathtext,mathtools,mathrsfs}
\usepackage{esint,bm,esvect,dsfont}
\usepackage{graphicx,epsfig,psfrag,xcolor,subcaption}
\usepackage{array,tabularx,tabulary,booktabs,multirow}
\graphicspath{{figures/}}
\usepackage{hyperref}
\usepackage{xcolor}

\newcommand{\figref}[1]{Fig.~\ref{#1}}

\def\ve{\varepsilon}

\def\tanh{\text{tanh}}

\def\ve{\varepsilon}

\def\th{\text{th}}
\def\sign{\text{sign}}

\def\eth{E_\textrm{Th}}
\def\ech{E_\textrm{ch}}

\begin{document}
\title{Instantons in the out-of-equilibrium Coulomb blockade}	
	\author{A.~S.~Dotdaev} 
		\affiliation{National University of Science and Technology MISIS, Moscow, 119049 Russia}
	\affiliation{Skolkovo Institute of Science and Technology, Moscow, 121205, Russia}
	\author{Ya.~Rodionov}
		\affiliation{Institute for Theoretical and Applied Electrodynamics, Moscow, 125412 Russia}
	\author{K.~Tikhonov}
	\affiliation{Skolkovo Institute of Science and Technology, Moscow, 121205, Russia}
	\affiliation{Landau Institute for Theoretical Physics, Russia}

\date{\today}

\begin{abstract}
Physical properties of single-electron devices in the week Coulomb blockade regime are significantly dependent on non-perturbative effects. They arise as instanton solutions of equations of motion for the corresponding Ambegaokar-Eckern-Schön action. In equilibrium those solutions are known as Korshunov instantons. In this paper we study non-equilibrium Ambegaokar-Eckern-Schön action using Keldysh technique. We found instantons for the most general stationary out-of-equilibrium state. We also found that action saddle-point value assumes a universal value irrespective of the stationary non-equilibrium state. 
\end{abstract}
\pacs{NaN}
\maketitle

\section{Introduction}
Single--electron devices have become an essential part of theoretical and experimental condensed matter physics\cite{grabert1991special,grabert2013single}. At low temperatures, electron transport through such devices is hindered by Coulomb blockade\cite{aleiner_review,schon_zaikin,likharev1985theory,averin1986coulomb,blanter2000phys,agkm} -- a phenomenon which remains a powerful tool for observation of interaction and quantum effects. Transport properties of single--electron devices 
are very sensitive to electric fields, making them useful in electrometry
and for providing current/temperature/resistance standards as well as in other applications\cite{likharev1999single}.
The related physics is also appealing from the theoretical point of view since strong Coulomb interactions often require a non-perturbative approach. Theoretical predictions for such systems are also well complemented with experimental studies, such as recent \cite{bitton_2011,bitton_2017}. 

\begin{figure}[t!]
	\centering
	\includegraphics[width=0.9\columnwidth]{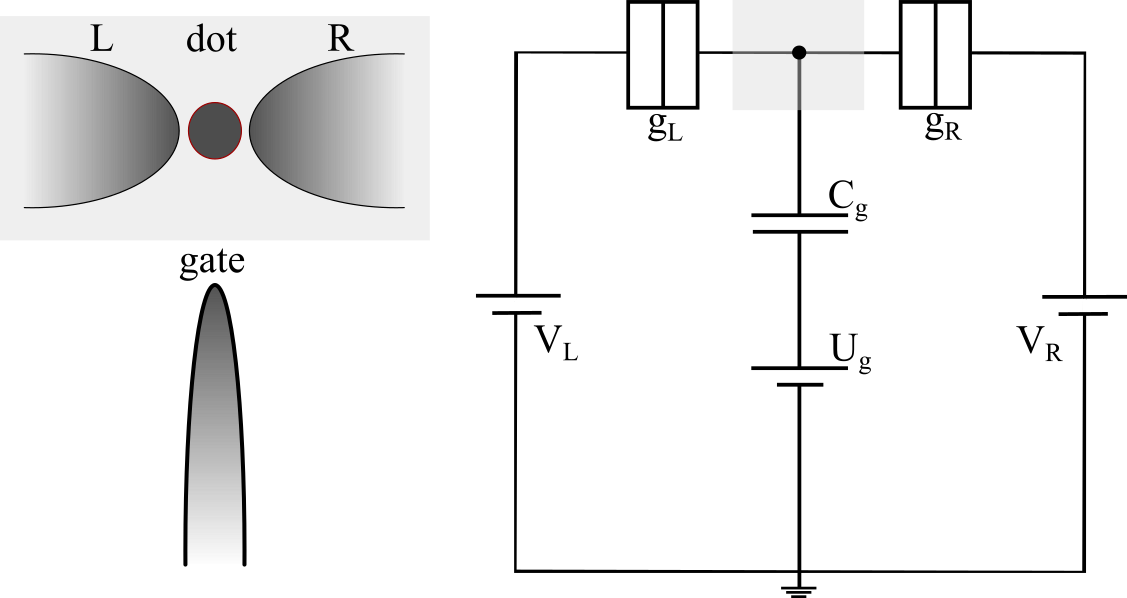}
	\caption{A schematic illustration of a single-electron transistor.}
	\label{fig:set}
\end{figure}

In this paper, we consider the setup of one of the most popular single-electron devices: a single-electron transistor (SET), \figref{fig:set}. It consists of an island, coupled to leads via tunnel junctions, characterized by dimensionless conductance $g_{R}$ ($g_{L}$) for right (left) lead. The Coloumb blockade regime is characterized by $T\ll \ech$ with $\ech=e^2/2C_0$, $C_0$ being electric capacitance of the dot. The island is also coupled to a gate electrode through a junction with capacitance $C_g$. The purpose of the gate electrode is in controlling the potential of electrons on the island by tuning the gate voltage $U_g$. 

The tunneling contacts are assumed to have large number of conductance channels $N_{\rm ch}\gg 1$ each of them having a small dimensionless tunneling conductance. The combination $g = g_l+g_R$ plays the role of the field coupling constant in our treatment and is assumed to be large $g\gg 1$. This, in turn, corresponds to the case of SET in the so called \textit{weak Coulomb blockade} regime.

Apart from the charging energy $\ech$, and temperature $T$ (in non-equilibrium conditions studied in this paper one may think of some effective temperature for the purpose of parametric estimates), the island is characterized by the following energy scales (see also Ref. \onlinecite{efetov_2003}): Thouless energy $\eth$ and the mean electron energy level spacing $\delta$. We study the regime $\eth \gg \ech \gg T \gg g\delta$. Thouless energy is the largest, allowing the island to be treated as a zero dimensional object and to be described by a single phase field $\phi(t)$. The parameter $\delta$ is considered to be small (\textit{metallic limit}). The lead is a reservoir of electrons which is large enough to be regarded as having continuous electron energy spectrum.
As a result, the condition $T \gg g\delta$ implies that electron coherent behaviour is supressed~\cite{efetov_tschersich}. 
In this regime, that is called \textit{interaction without coherence}\cite{agkm}, the system is well described by Ambe\-ga\-okar-Eckern-Schon action (AES)\cite{aes}, that is also referred to as dissipative action (it is also known to describe a quantum particle on a ring\cite{guinea2002aharonov,guinea2003low}).

For $g \gg 1$ the effect of the Coulomb blockade in the conductance of a SET shows up as weak (exponentially small in $g$) oscillations with $q=C_g U_g/e$, Ref. \onlinecite{agkm}.
Exponential smallness implies that formally these oscillations are produced by instantons -- nontrivial solutions of equations of motion of AES action. The linear response of SET under thermal equilibrium is described by analytically continued imaginary--time response functions. The instantons of Euclidian AES action were found by Korshunov\cite{korshunov}, and their contribution to observables has been thoroughly studied \cite{agkm,rodionov_2010} in the framework of Matsubara formalism. Although this approach allows to evaluate non--linear response functions under assumption of relatively strong electron--electron interactions, establishing thermal equilibrium on the islands at the temperature of the leads\cite{Igor2016}, it is inapplicable in a more general non--equilibrium situation.

 The first attempt to tackle out-of-equilibrium situation was undertaken recently by Titov and Gutman\cite{titov_2016}. In the framework of real--time Keldysh technique, they considered a particular case of a SET, biased by finite voltage. They generalized Korshunov instantons to this particular case and found that saddle--point action, determining the order of  magnitude of the effect, is independent on the voltage bias and remains the same as in equilibrium. 
 
 In the present paper, we construct instantons of AES action in a general stationary non--equilibrium situation and show that the value of the action on instantons is independent on distribution functions of electrons on the leads and on the island. Hence, the main exponent $e^{iS}$, computed on instanton configuration is given by a universal factor $e^{-g/2}$ (as in equilibrium), independent of particular realization of out--of--equilibrium regime. 
 
 The paper is organized as follows. In the Section II, we formulate the model and the effective AES action formalism.
 In the Section III, we solve AES equations of motion and present exact instanton solutions for the case of the most general stationary non-equilibrium situation.
 In the  Section IV we discuss the implications of the obtained results.
 
\section{Formalism}
The Hamiltonian for a SET is as follows:be
\begin{equation}
\hat{H}=\hat{H}_0+\hat{H}_t+\hat{H}_{\text{ch}}.
\end{equation}
It consists of following contributions. The Hamiltonian of free electrons on the island and the leads,
\begin{equation}
\hat{H}_0=\sum\limits_{k, \alpha}\ve_k^{(\alpha)} \hat{a}_k^{(\alpha)\dagger}\hat{a}_k^{(\alpha)} + \sum\limits_{m}\ve_{m}^{(d)} \hat{d}_{m}^{\dagger}\hat{d}_{m},
\end{equation}
where $\hat{a}_k^{(\alpha)\dagger}$$(\hat{d}_{m}^{\dagger})$ is the creation operator of an electron with energy $\ve_k^{(\alpha)}$$(\ve_{m}^{(d)})$ on a lead (the island), the index $\alpha=R,L$ corresponds for right and left leads. The tunnelling Hamiltonian,
\begin{equation}
\hat{H}_t=\sum\limits_{k,m} \Big(t_{km}^{R}\hat{a}_k^{R\dagger}\hat{d}_{m}
+ t_{km}^{L}\hat{a}_k^{L\dagger}\hat{d}_{m} + \text{h.c.} \Big),
\end{equation}
where $t_{km}^{R,L}$ are amplitudes of tunnelling between the right (left) lead and the island. The charging Hamiltonian,
\begin{equation}
\hat{H}_{\text{ch}}={E}_{\text{ch}}(\sum\limits_{m}\hat{d}_{m}^{\dagger}\hat{d}_{m} - q)^2,
\end{equation}
where ${E}_{\text{ch}}$ is the charging energy and $q$ is a background charge on the island set by gate potential.

The total action for the SET is given by the following integral along Keldysh contour:
\begin{gather}
S=\int_K \! dt \left(i\hat{a}^{\dagger}\partial_t\hat{a}+i\hat{d}^{\dagger}\partial_t\hat{d} -\hat{H} \right)
\end{gather}
Integration over fermion fields $a, d$ with subsequent $1/N_{\rm ch}$ expansion in the number of conductance channels leads to the well-known AES action $S = S_{\rm ch}+S_{d}$ where
\begin{gather}
S_{\text{ch}}=\int_K \Bigg( \frac{\dot\phi^2(t)}{2\ech} -q\dot\phi(t)\Bigg)dt
\end{gather}
is the charging action and 

\begin{equation} \label{action}
S_d = \frac{ig}{4} \iint \text{d}t\text{d}t' \bar{\mathrm{X}}^T(t) \hat{\Pi}(t,t') \mathrm{X}(t')
\end{equation}
is the dissipative part of the action.
Here, $g=g_L+g_R$ and
\begin{equation} \begin{gathered}
\hat{\Pi}=
  \left( {\begin{array}{cc}
   \Pi^K + \Pi^R + \Pi^A & \; -\Pi^K + \Pi^R - \Pi^A \\
   -\Pi^K - \Pi^R + \Pi^A & \; \Pi^K - \Pi^R - \Pi^A \\
  \end{array} } \right)
\end{gathered}\end{equation}
is the polarization operator, its components:
\begin{equation}\label{eq:general_formula_for_KRA} \begin{gathered}
\Pi^{R,A}_{\omega}=\mp i\sum\limits_{\alpha=R,L}\frac{g_\alpha}{g} \int[F^d_\ve-F^\alpha_{\ve-\omega}]\frac{d\ve}{2\pi},\\
\Pi^{K}_{\omega}=2i\sum\limits_{\alpha=R,L}\frac{g_\alpha}{g}\int[1-F^d_\ve F^\alpha_{\ve-\omega}]\frac{d\ve}{2\pi}.
\end{gathered}\end{equation}
The field functions,
\begin{equation} \begin{gathered}
\bar{\mathrm{X}}^T(t) = \Big( \frac{1}{\chi_{+}(t)} \quad \frac{1}{\chi_{-}(t)} \Big), \qquad
\mathrm{X}(t) = \left( {\begin{array}{c}
   \chi_{+}(t)\\
   \chi_{-}(t) \\
  \end{array} } \right),
\\ \chi_{\pm}(t)=e^{-i \phi_{\pm}(t)}
\end{gathered}\end{equation}
are vectors in Keldysh space. The $\pm$ index of $\phi$ corresponds to the phase field $\phi(t)$ on the upper (lower) branch of Keldysh contour\cite{keldysh_1965}. Functions 
$F^{d(L,R)}$ represent electron distribution functions on the island (leads).

\section{Instantons}
\subsection{Regimes of non--equilibrium}
Depending on the size of the island of a SET, different regimes of out--of --equilibrium island electorns can be realized. The regimes are defined by two competing electron relaxation rates:   $\tau^{-1}_{\textrm{ee}}$ (relaxation due to electron-electron interactions),
$\tau^{-1}_{E}$ (relaxation due to tunneling to reservoirs). We discard possible electron--phonon interaction as it gets frozen out at typically low experimental temperatures.
The corresponding electron--electron relaxation rates for mesoscopic devices were derived in Refs. \onlinecite{blanter,imry}. For transparency one may  articulate two distinct non--equilibrium limiting situations. i) The \textit{quasi-equilibrium} regime, when  e--e relaxation prevails over relaxation due to tunneling $(\tau^{-1}_{\textrm{ee}}\gg \tau^{-1}_{E})$. In this case the electrons are thermalized to
\begin{gather}
\label{temp_noneq}
    T_{d}=\frac{g_L T_L+g_R T_R}{g},    
\end{gather}
determined by the respective temperatures of the reservoirs. ii) The 
\textit{non-equilibrium} regime when the thermalization is governed by the tunneling events. In the latter case the non-fermi electron distribution function sets up on an island: 
\begin{equation}
F^{d}=\frac{g_L}{g} F^L+\frac{g_R}{g} F^R.
\end{equation}

According to the estimates made in Ref~\onlinecite{rodionov_2010}, both scenarios are experimentally relevant. For simplicity, we address the quasi--equilibrium situation at first and then turn to the more general case.

\subsection{Quasi--equilibrium scenario}
In this case the equilibrium is upset by a finite difference of the temperatures of electrons in the leads.  The temperatures of the leads are assumed to be constant in time.
The island electrons are then thermalized to a stationary non--equilibrium temperature~\eqref{temp_noneq}.


The equations of motion for this particular case are:
\begin{equation}\label{motion_special}
\begin{gathered}
\chi_{\pm}\frac{\delta S_d}{\delta\chi_{\pm}}=
\frac{1}{2}\int(s_++s_-)
    \left(\frac{\chi_{\pm}}{\chi_{\pm}'}-\frac{\chi_{\pm}'}{\chi_{\pm}}\right)dt'\\
	+\int s_{\mp}\frac{\chi_{\mp}'}{\chi_{\pm}}dt'-
    \int s_{\mp}\frac{\chi_{\pm}}{\chi_{\mp}'}dt'=0,\\
\end{gathered}\end{equation}
where $s_{\pm} \equiv s_{\pm}(t-t')$ are calculated in Appendix \ref{app:special}; $\chi_{\pm}' \equiv \chi_{\pm}(t')$, $\chi_{\pm} \equiv \chi_{\pm}(t)$.
The equations of motion \eqref{motion_special} are nonlinear integral equations with a singular kernel; general approaches for such equations are not developed or at least rather complicated. However, they may be solved explicitly once appropriate assumptions about analytical properties of solutions are made.

\begin{figure}[t!]
	\centering
	\includegraphics[width=0.8\columnwidth]{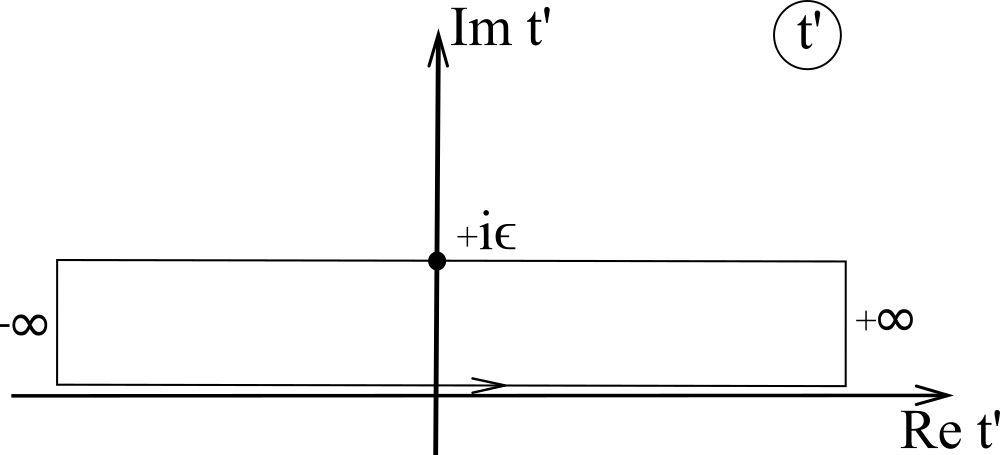}
	\caption{The contour of integration in the plane of complex $t'$; $\epsilon$ is a finite positive number, and so is $\pi-\epsilon$.}
	\label{fig:contour}
\end{figure}

In spirit of Refs. \onlinecite{braunecker2006response,muzykantskii2003scattering}, we combine the fields $\chi_{\pm}^{-1}(t)$ into a single function of complex $t$:
\begin{equation}
\frac{1}{\chi_{\pm}(t)}=\frac{1}{\chi(t \mp i\delta)}, \; \delta>0.
\end{equation}
Inspired by the solution, found in Ref. \onlinecite{titov_2016}, we assume that the function $\chi^{-1}(t)$ has a pole at $t=t_0$ (with real $t_0$) and has no zeros or branch cuts on the real axis. Then $\chi(t)$ is analytical in some vicinity of real axis, so  effectively $\chi_{+}(t)$ = $\chi_{-}(t)$ for all real $t$. The integrals in the equations of motion are expressed as contour integrals taken on the contour shown in the \figref{fig:contour} and can be calculated via residues.
The Eqs.~\eqref{motion_special} are then reduced to single differential equation:

\begin{equation}
\frac{1}{\pi^2}\frac{d}{dt}\!\left(\frac{1}{\chi_{\pm}(t)} \right) =\frac{C_{-1}}{\ech}  s_{\pm}(t_0\!-\!t), \end{equation}
where $C_{-1}/\ech$ is the residue of $1/\chi(t)$ in the point $t_0$. In this notation, $C_{-1}$ is dimensionless and the dimensional factor $\ech$ is natural from the point of view of subsequent computations.

The instanton trajectory can be found explicitely:
\begin{equation}\label{instanton_special}
\frac{1}{\chi_{\pm}(t)} = C_{-1}\Big(\frac{1}{\ech}\int\limits_{-\infty}^{t} dt' s_{\pm}(t_0\!-\!t')+ B\Big).
\end{equation}

Here, we note that the freedom of choice of $t_0$ reflects the time translational invariance of AES action. Additionally,  $B$  is a parameter such that $\chi^{-1}(t)$ does not have zeroes on a real axis.This implies that $B$ is real and belongs to the interval $[-\ech^{-1}\mathrm{Re} \int_{-\infty}^{\infty}dt s_{+}(t),0]$.


 In equilibrium ($T_L=T_R$) ~\eqref{instanton_special} become Korshunov instantons: $\chi_\pm(t)\sim \coth(t-t_0)$, Ref.~\onlinecite{titov_2016}.
Note that the simple pole structure of saddle--point stationary configurations of real--time path integrals is typical in more elementary (non-dissipative) quantum mechanical problems, such as tunneling in the double--well potential\cite{cherman2014real}. 

Parameters $B$ and $t_0$ are two zero modes of the fluctuation action (expanded near instanton solution). Their range defines preexponential factor in the computation of all observables.
However, the computation of physical observables is beyond the scope of this paper.

The action on these instantons is easily computed (using integrals along the contours as shown in \figref{fig:contour}) and reads
\begin{gather}
\begin{split}
S_0&= \frac{ig}{2},\\
 S_{\rm ch}&= \pi\Big(\int\limits_{0}^\infty \Big[s_{+}(t)+\frac{i}{\pi^2 t^2}\Big]dt+B\Big).
    \end{split}
\end{gather} 
Here, $S_0,\ S_{\rm ch}$ are dissipative and charging parts respectively.
Quite surprisingly, the dissipatice part $S_0$ is the same as in equilibrium, despite the system being essentially out-of-equilibrium, with instantons and the action  depending on each of the temperatures $T_L$ and $T_R$. A similar striking feature is present in the paper \cite{titov_2016}. For that reason we decided to study a system in a state of quite general stationary non-equilibrium. In our case that means that the electrons on the island and the leads have arbitrary electron distrubution functions.

\subsection{General non-equilibrium}
Now we consider a case of stationary arbitrary non-equilibrium. 
The components of polarization operator
depend  on time difference $(t-t')$ only (due to stationariness of the distribution functions).
Equations of motion in this case are
\begin{equation}  \label{eq:motion_general}
\begin{gathered}
\int \Pi_{\pm\pm}(t'\!-\!t) \Big(\frac{\chi_{\pm}}{\chi_{\pm}'} -\frac{\chi_{\pm}'}{\chi_{\pm}}\Big) dt'\\
+\int \Pi_{\mp\pm}(t'\!-\!t) \frac{\chi_{\pm}}{\chi_{\mp}'} dt' 
-\int \Pi_{\pm\mp}(t\!-\!t') \frac{\chi_{\mp}'}{\chi_{\pm}}dt'=0.
\end{gathered}\end{equation}
Here $\Pi_{++}$ and $\Pi_{--}, \quad \Pi_{-+}$ and $\Pi_{+-}$ are the matrix elements of $\hat\Pi$ (Appendix \ref{app:general}). 

What is important is that despite $\widehat{\Pi}$ consisting of distribution functions of arbitrary non-equilibrium,  its Laurent expansion in the vicinity of $t-t^\prime=0$ is universal and almost independent of distribution functions. For example, the crucial Keldysh component reads: \begin{gather}
    \Pi^{K}(t)=-\frac{i}{2\pi^2}\left( \frac{1}{(t+i0)^2} + \frac{1}{(t-i0)^2} \right) +\Pi_{\rm reg}(t),
\end{gather}
where $\Pi_{\rm reg}(t)$ is a function analytical on the real axis, see the derivation in Appendix B.

The equations of motion now become:
\begin{equation}\label{eq:de}
\begin{gathered}
\mp\frac{2i}{\pi^2} \frac{\dot{\chi}_{\pm}(t)}{\chi_{\pm}(t)}
+ 
\frac{2i}{\pi^2} \frac{C_{-1}}{(t -t_0 \mp i0)^2} \chi_{\pm}(t)
\\
\pm 2\pi iC_{-1} \chi_{\pm}(t) \Pi_{\text{reg}}(t_0-t) =0.\\
\end{gathered}
\end{equation}
Their solutions are:
\begin{equation}\label{eq:general_equation_of_motion}
\begin{split}
\frac{1}{\chi_{\pm}(t)} =C_{-1}\Big(\frac{1}{E_{\rm ch}}\frac{1}{t-t_0 \mp i0} + \chi_{\text{reg}}(t-t_0)\Big),
\end{split}  \end{equation}
where $\chi_{\text{reg}}(t) = E_{\rm ch}^{-1}\int_{-\infty}^t \Pi_{\text{reg}}(-t') dt'+B$ is a function which is regular on the real axis, $B$ is complex parameter bonded by the condition that $\chi_{\pm}^{-1}(t)$ does not have real zeros.

The saddle point value of the dissipative and the charging action, calculated on these instantons  read respectively
\begin{gather}
    \begin{split}
    \label{action-non}
S_0&= \frac{ig}{2},\\
S_{\rm ch}&= \pi\Big(\int\limits_{0}^\infty \Pi_{\rm reg}(t)dt+B\Big).
    \end{split}
\end{gather}
And we stress that the dissipative term is again the same as in equilibrium.

Multiinstanton solutions are given by products of different single-instantons:
\begin{equation}\label{}
\frac{1}{\chi_{W\pm}(t)} =\frac{C_{-1}}{E^W_{\rm ch}}\prod\limits_{i=1}^{W} \left( \frac{1}{t-t_i \mp i0} + \chi_{i\,\text{reg}}(t-t_i)\right)\end{equation}
The dissipative action on these trajectories is $WS_0$.

Multiinstanton contribution corresponds to harmonics of conductance vs. gate voltage dependence, observed in Ref. \onlinecite{bitton_2017}. Equations \eqref{eq:general_equation_of_motion} and ~\eqref{action-non} are the main results of the paper.

We thus have proven that the saddle-point value of dissipative acton is the same for any out-of-equilibrium system (as long as the distribution functions are stationary).

\section{conslusion}
We have studied non-perturbative 
solutions of AES action describing single-electron device using Keldysh technique. We addressed arbitrary non-equilibrium regime and obtained the exact form of
instantons. We calculated the
saddle-point value of the action. Strikingly, it does not depend on the electron distribution function as long as it is stationary. From our point of view this fact points to some hidden symmetry in this problem, which has not been  established yet.

The instanton action determines the leading exponential dependence $e^{-g/2}$ of related physical observables (e.g. renormalized charging energy, SET conductance \textit{etc}). The dependence on the electron distribution will show up in the preexponential factor (the computation would require account for Gaussian fluctuations around instantons together with integration over the zero modes $t_0$ and $B$). It would also be interesting to develop this approach for non-stationary out--of--equlibrium setup.
\section{Acknowledgments}

We thank I. Burmistrov, I. Gornyi and A. G. Semenov for useful discussions.

Ya.I.R. acknowledges the support of the Russian Foundation for Basic Research, Projects No. 19-02-00421, JSPS-RFBR Grants No. 19-52-50015 and
No. JPJSBP120194828 and by Deutsche Forschungsgemeinschaft (DFG), grant No. EV 30/14-1 and
of the Ministry of Science and Higher Education of the Russian Federation in the
framework of Increase Competitiveness Program of NUST MISiS, Grant No. K2-2020-038.

\begin{widetext}
\input{appendix}

\end{widetext}
\bibliography{instanton}

\end{document}

%% file: appendix.tex
\appendix
\section{Polarization operator for special non-equilibrium}\label{app:special}
Here we calculate the components of polarization operator for the case of electrons on the dot and on the lead having different temperatures.
Electrons on the dot and the lead obey Fermi distribution, its Wigner transform is $F(\ve)=\th (\ve/2T)$. Substituting it to the expressions \eqref{eq:general_formula_for_KRA} we get for the retarded (R) and the advanced (A) components (in the frequency domain and the time domain):

\begin{equation}
\Pi^{R,A}(\omega) = \mp \frac{i \omega}{\pi}, 
\qquad 
\Pi^{R,A}(t)= \int e^{-i\omega t} \Pi^{R,A}(\omega) \frac{d\omega}{2 \pi} = \pm \frac{1}{\pi} \delta' (t).
\end{equation}
Keldysh component is calcuated right away in the time domain:

\begin{equation}\begin{gathered}
\Pi^K(t) =2i\sum\limits_{\alpha=L,R}\frac{g_{\alpha}}{g}\int e^{-it\omega} \frac{d\omega }{2\pi} \int(1-\tanh\frac{\ve}{2T_d}\tanh\frac{\ve-\omega}{2 T_{\alpha}}) \frac{d\ve}{2\pi}
\\
=2i\sum\limits_{\alpha=L,R}\frac{g_{\alpha}}{g}\int e^{-it \ve}\frac{d\ve}{2 \pi}  \int e^{-it(\omega-\ve)} \frac{d(\omega-\ve)}{2\pi}  (1-\th \frac{\ve}{2T_d} \th \frac{\ve-\omega}{2 T_{\alpha}})\\
= -i \sum\limits_{\alpha=L,R}\frac{g_{\alpha}}{g} \text{Re} \Big( \frac{T_{\alpha}}{\sinh (\pi T_{\alpha} t + i0)} \frac{T_d}{\sinh(\pi T_d t + i0)}  \Big)
=-\frac{i}{2}\Big(s_+(t)+s_-(t) \Big).
\end{gathered}  \end{equation}
Here the notation is introduced:
$$
s_{\pm} (t) \equiv \frac{g_R}{g}\frac{T_R}{\sinh (\pi T_R t \pm i0)} \frac{T_d}{\sinh(\pi T_d t \pm i0)}
+\frac{g_L}{g}\frac{T_L}{\sinh (\pi T_L t \pm i0)} \frac{T_d}{\sinh(\pi T_d t \pm i0)}.
$$
The derivative of delta function is represented as follows: 
$
2\pi\delta'(t) = 4\pi^2 i\big(s_+(t) - s_-(t) \big),
$
so
\begin{equation}
\Pi^{R}(t)= \frac{i}{2}\Big(s_+(t) - s_-(t)\Big).
\end{equation}
The notations for matrix elements of the polarization operator \eqref{eq:general_formula_for_KRA},
$$
\hat{\Pi} \; : \;
  \left( {\begin{array}{cc}
   \Pi_{++} & \Pi_{+-}  \\
   \Pi_{-+} & \Pi_{--}  \\
  \end{array} }. \right)
$$
The elements of the polarization operator in time domain:
\begin{gather}
\Pi_{++} =-\frac{i}{2}\Big( s_+ + s_- \Big), \qquad
\Pi_{+-} =is_-,\qquad
\Pi_{-+} =is_+.
\end{gather}

\section{General non-equilibrium}\label{app:general}
Here we discuss the analytical properties of polarization operator in a general case, in which $F^d$ and $F^{R,L}$ are any physically meaningful stationary electron distribution functions.
The retarded and the advanced components of polarization operator (from \eqref{eq:general_formula_for_KRA}),
\begin{gather}
    \Pi^{R,A}(\omega)=\mp i\frac{2\omega+\mu_0}{2\pi},\quad 
    \mu_0 \equiv \sum\limits_{\alpha=L,R}\frac{g_{\alpha}}{g}\int[F^d_\ve-F^{\alpha}_{\ve}]d\ve.
\end{gather}
In the time domain,
\begin{gather}
    \Pi^{R,A}(t)=\pm \frac{1}{\pi}\delta^\prime(t)\mp\frac{i\mu_0}{2\pi}\delta(t).
\end{gather}
Here, the value of the functional $\mu_0$ reflects assymetry of distribution functions; for Fermi distributions it is equal to zero. 

Keldysh part is less trivial.
Let us represent the non-equilibrium functions as $F_\ve=\sign(\ve/2)+\Delta F_\ve$, where $\Delta F_\ve\rightarrow0\big|_{\ve\rightarrow\pm\infty}$ is the asymmetric part of the distribution function. Let us derive a general formula for Keldysh component:

\begin{equation*} \begin{gathered}
\Pi^K(\omega)=2i\sum\limits_{\alpha=L,R}\frac{g_{\alpha}}{g}\int[1-F^d_\ve F^{\alpha}_{\ve-\omega}]\frac{d\ve}{2\pi}
=2i\int(1-\sign(\frac{\ve}{2})\sign(\frac{\ve-\omega}{2}))\frac{d\ve}{2\pi}
\\
-2i\int\frac{d\ve}{2\pi}\sum\limits_{\alpha=L,R}\frac{g_{\alpha}}{g}\big(\Delta F^d_\ve\sign(\frac{\ve-\omega}{2})+\Delta F^{\alpha}_{\ve-\omega}\sign(\frac{\ve}{2})\big)
-2i\int\frac{d\ve}{2\pi}\sum\limits_{\alpha=L,R}\frac{g_{\alpha}}{g}\Delta F^d_\ve \Delta F^{\alpha}_{\ve-\omega} 
\quad  \equiv I(\omega)+II(\omega)+II(\omega).
\end{gathered}\end{equation*}
In the time domain, the integral $I$ gives the second-order poles in the points $t=\pm i0$:
$$
2i \int\frac{d\omega}{2\pi}\int(1-\sign(\frac{\ve}{2})\sign(\frac{\ve-\omega}{2}))\frac{d\ve}{2\pi}e^{-i\omega t}
=\frac{i}{2 \pi^2}\int d\omega |\omega| e^{-i\omega t}
=- \frac{i}{2\pi^2 (t+i0)^2} - \frac{i}{2\pi^2 (t-i0)^2}.
$$
The corrections $\Delta F_{\ve}$ to distribution functions are assumed to decrease fast enough so that Fourier transforms of $II(\omega)$, $III(\omega)$ are analytical on real axis of $t$. According to that, the matrix elements of the integral operator,

\begin{equation*} \begin{gathered}
\Pi_{++} = \Pi_{--}= \Pi^K+\Pi^R+\Pi^A = \Pi^K,\\
\Pi_{+-}(t) = -\Pi^K(t)+\Pi^R(t)-\Pi^A(t) 
=\frac{i}{\pi^2} \frac{1}{(t-i0)^2} - \frac{i\mu_0}{\pi}\delta(t) -\Pi_{\rm reg}(t) ,\\
\Pi_{-+}(t) = -\Pi^K(t)-\Pi^R(t)+\Pi^A(t) =
\frac{i}{\pi^2} \frac{1}{(t+i0)^2} + \frac{i\mu_0}{\pi}\delta(t) -\Pi_{\rm reg}(t).
\end{gathered}\end{equation*}
With the use of a notation
$$
p_{\pm}(t) \equiv \frac{i}{\pi^2} \frac{1}{(t \pm i0)^2} \pm  \frac{i\mu_0}{\pi}\delta(t)
$$
the operator $\hat\Pi$ is conveniently expressed:
\begin{equation} \begin{gathered}
\hat{\Pi}(t)= 
\left( {\begin{array}{cc}
    -\frac{1}{2}(p_+ + p_-) &  p_- \\
   p_+ & -\frac{1}{2}(p_+ + p_-)\\
\end{array} } \right)_t
+
\Pi^K_{\rm reg}(t) 
    \left( {\begin{array}{cc}
   1 & -1\\
   -1 & 1\\
    \end{array} } \right).
\end{gathered}\end{equation}
The parameter $\mu_0$ is responsible for presence of first-order poles in the points $t=\pm i0$. For the case of electrons obeying Fermi distribution, as we have shown, there are only second-order poles at $t=\pm i0$. A first-order pole, however, does not affect either the analytical structure of instantons or the saddle-point value of dissipative action.

The l.h.ss. of the equations \eqref{eq:motion_general} are split to  singular parts,
$$
\frac{1}{2} \int (p_+ + p_-)_{(t,t')} \left( \frac{\chi_{\pm}(t)}{\chi_{\pm}(t')} - \frac{\chi_{\pm}(t')}{\chi_{\pm}(t)} \right) dt'
- \int p_- (t,t') \left( \frac{\chi_{\mp}(t')}{\chi_{\pm}(t)}  - \frac{\chi_{\pm}(t)}{\chi_{\mp}(t')} \right) dt' ,
$$
and regular parts,
\begin{equation*}\label{Preg} \begin{gathered}
\int \Pi_{\text{reg}}(t',t) \left( \frac{\chi_{\pm}(t)}{\chi_{\pm}(t')} - \frac{\chi_{\pm}(t)}{\chi_{\mp}(t')} \right) dt'
+\int \Pi_{\text{reg}}(t,t') \left( \frac{\chi_{\mp}(t')}{\chi_{\pm}(t)} - \frac{\chi_{\pm}(t')}{\chi_{\pm}(t)} \right) dt'.
\end{gathered}\end{equation*}
If the field functions $\chi_{\pm}$ satisfy the assumptions made in the main text, these integrals are easily calculated and
the equations \eqref{eq:motion_general} are reduced to simple differential equations \eqref{eq:de}.